\documentclass[runningheads,openany]{svmult}
\usepackage[T1]{fontenc}
\usepackage{palatino}
\usepackage{euler}
\usepackage{graphicx}  
\usepackage{physprbb}  
\usepackage{bg2018}  

\begin{document}
\title*{The Roper resonance as a meson-baryon molecular state}

\author{B. Golli}
\authorrunning{B. Golli}
\institute{
Faculty of Education,
              University of Ljubljana
 and
Jo\v{z}ef Stefan Institute, 
              1000 Ljubljana, Slovenia}

\titlerunning{The Roper resonance as a meson-baryon molecular state}

\maketitle

\begin{abstract}
The recently proposed mechanism for the formation of the Roper
resonance, in which a dynamically generated state as well as
a genuine three-quark resonant state play an equally important
role, is confronted with the model proposed almost twenty years
ago in which the Roper is pictured as a molecular state of the
nucleon and the $\sigma$ meson.
\end{abstract}

\noindent
Our recent investigation on the nature of the Roper resonance~\cite{PRC} 
has been motivated by the results of lattice QCD simulation in 
the P11 partial wave by the Graz-Ljubljana and the Adelaide 
groups \cite{lang16,kiratidis17} 
that have included beside three-quark interpolating fields also
operators for $\pi N$  in relative $p$-wave and  $\sigma N$ in
$s$-wave, and have found no evidence for a dominant three-quark
configuration below 1.65~GeV.
In our research we use a coupled channel approach 
which has been previously successfully applied to describe 
meson scattering and photo- and electro-production 
in several partial waves in the intermediate energy region~\cite{%
EPJ2005,EPJ2008,EPJ2009,EPJ2011,EPJ2013,EPJ2016}.
In the present analysis of the Roper resonance we include
the $\pi N$, $\pi\Delta$, and $\sigma N$ channels and solve 
the Lippmann-Schwinger equation for the meson amplitudes to 
all orders in the approximation of a separable kernel.
We have concluded that while
the mass of the resonance is determined by the dynamically 
generated state, an admixture of the (1s)$^2$(2s)$^1$ component 
at an energy around 2~GeV turns out to be crucial to reproduce 
the experimental width and the modulus of the resonance pole.
The mass of the dynamically generated state appears
typically 100~MeV below the (nominal) nucleon-sigma threshold.
This result agrees well with the prediction of a completely
different approach that we studied in the 2001 paper~\cite{Pedro}
in which we discussed the possibility that the Roper was 
a molecular state of the nucleon and the $\sigma$ meson.
In the following we review the main features of this
molecular state and its relation to the dynamically
generated state emerging in the coupled channel approach.

\begin{figure}[htbp]
\begin{center}
\hbox to \hsize{\kern-4mm\includegraphics[width=64mm]{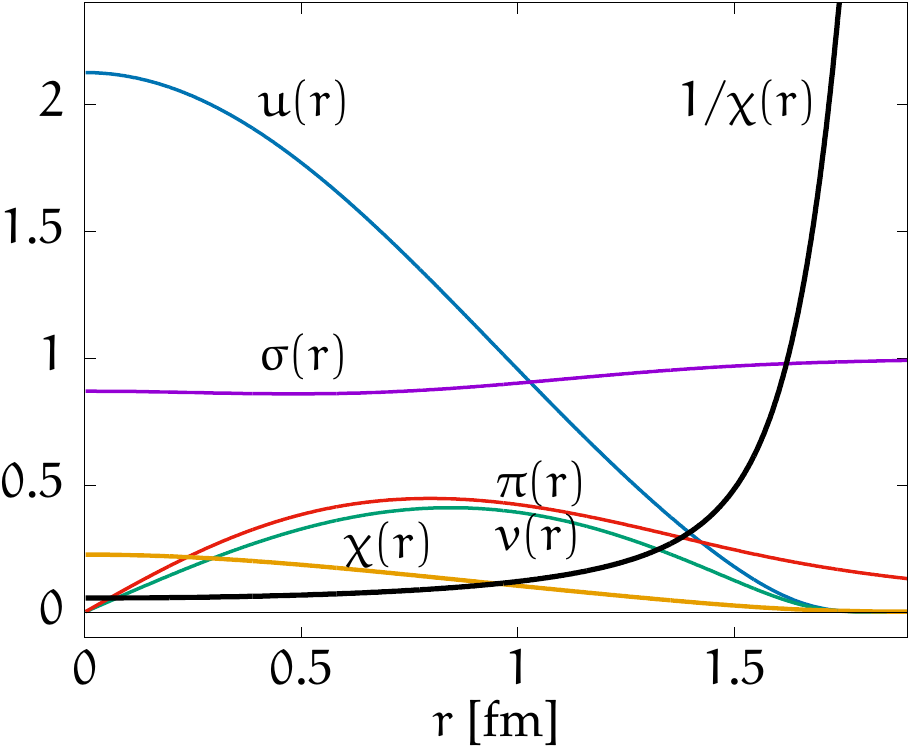}
\includegraphics[width=65mm]{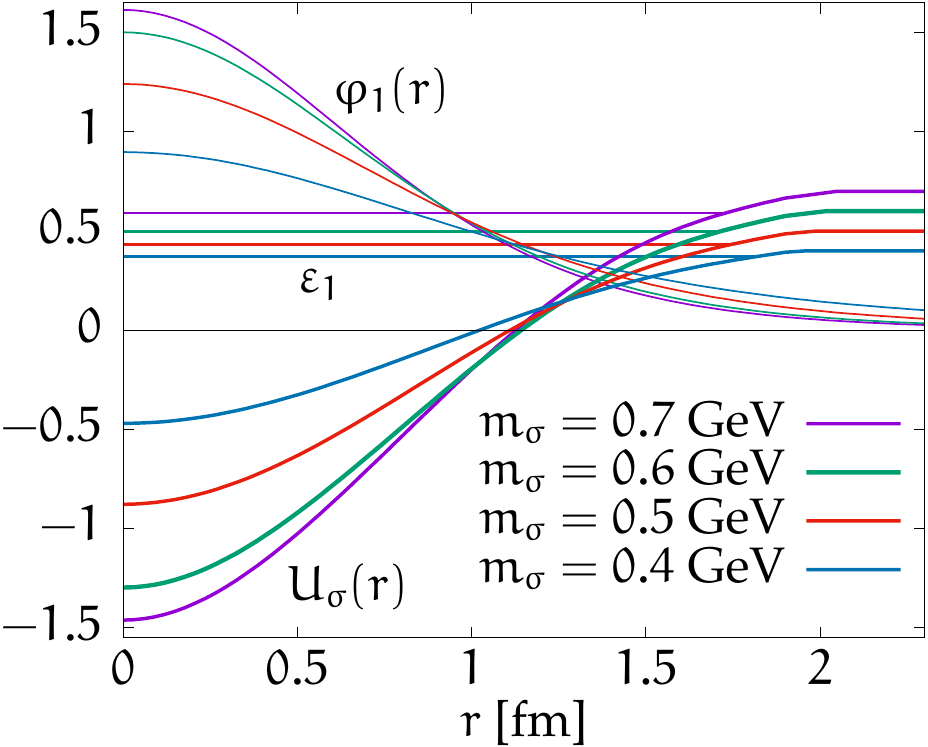}\kern-2mm}\hfill
\centerline{a)\hspace{63mm} b) \hspace{7mm}}
\vspace{-12pt}
\vspace{3mm} \caption{\label{fig:sc}
a) Self-consistently determined quark and boson
(in units of $f_\pi$) fields in the CDM.
b) Effective potential for the $\sigma$ meson and the lowest 
eigenvalue $\varepsilon_1$ of the corresponding Klein-Gordon 
equation (in units of GeV) for different choices of the $\sigma$ mass.}
\end{center}
\end{figure}

In our early approaches to describe the nucleon and the
$\Delta(1232)$ we used a chiral version of the 
linear $\sigma$-model with quarks  and determined the quark 
and meson fields self-consistenly.
This model does not work for higher nucleon excitations
since the energy of the excited quark turns out to be higher 
than the free quark mass.  
In order to ensure confining we used in~\cite{Pedro} a chiral 
version of the Cromodielectric model which included, beside the 
$\sigma$ and the pion fields, the chromodielectric field $\chi$.
The coupling of the $\chi$ field to the quark and meson fields
is taken in the form:
\begin{equation}
  \mathcal{L}_{\mathrm{int}} =
{g\over{\chi}}\, \bar{q}
    (\hat{\sigma}+\mathrm{i}\pol{\tau}\cdot\hat{\pol{\pi}}\gamma_5)q\,,
\label{Lint}
\end{equation}
such that for $r\to\infty$, ${\chi}(r)\to0$, while the
quark mass in this limit behaves as
$$
    m_q = {g\sigma(r)\over{\chi}(r)}= {gf_\pi\over{\chi}(r)}\to \infty\,,
$$
which means that the quarks are bound.
A typical self-consistent solution for the fields is shown 
in Fig.~\ref{fig:sc} a).

We next expanded the field operators of the bosons around  their 
expectation values in the ground state $|{\rm N}\rangle$;
the $\sigma$ operator can be written as:
$$
   \hat{\sigma}(\vec{r}) =
       \sum_n{1\over\sqrt{2\varepsilon_n}}\, \varphi_n(r)
        {1\over\sqrt{4\pi}}
        \left[\tilde{a}_n + \tilde{a}^\dagger_n\right] + \sigma(r)\,,
\qquad
  \tilde{a}_n|N\rangle = 0\,.
$$
The stability conditions implies a Klein-Gordon equation 
for the $\sigma$-meson modes:
$$
 \left(-\nabla^2 + m_\sigma^2 
     + U_\sigma(r)\right)\varphi_n(r)
  = \varepsilon_n^2 \varphi_n(r)\;,
\qquad
   U_\sigma(r) =  {{\rm d}^2 V(\sigma(r))\over{\rm d}\sigma(r)^2}\,.
$$
Here $V$ stands for the potential originating from (\ref{Lint}) 
and the potential parts of the $\sigma$-model.
The potential $U_\sigma$ (see Fig.~\ref{fig:sc}b)) is attractive 
and supports a bound state which can be interpreted as 
a molecular state of the nucleon and (one quantum of) the $\sigma$.
The corresponding potential for the $\chi$ field turns out to be 
repulsive, which means that the model does not predict glueball states.

In \cite{Pedro} this excitation of the $\sigma$ field was confronted 
with the excitation
of the quark core in which one quark wass promoted to the $2s$ orbit.
In the self-consistent solution the $2s-1s$ energy splitting
turned out to be smaller than the corresponding vibrational energy 
$\epsilon_1$, and the conclusion of our work  was that the Roper 
consisted of the dominant quark excitation and a $\sim 10$~\%
admixture of the molecular state.
However, in that work we used -- in accordance with then accepted
values -- a relatively large $\sigma$ mass between 0.7~GeV and 1.2~GeV.
With the present value $\sim 0.5$~GeV, the lowest eigenmode 
$\varepsilon_1$ decreases (see  Fig.~\ref{fig:sc}b)), while,
assuming a somewhat smaller nucleon size, the $2s-1s$ splitting
increases, such that the molecular state may eventually become
the dominant component of the Roper resonance.

In our recent paper \cite{PRC} we study the formation of
the resonance in this partial wave in a coupled-channel approach
including the $\pi N$, $\pi\Delta$ and $\sigma N$  channels.
The Cloudy Bag Model is used to fix the quark-pion vertices while 
the $s$-wave $\sigma$-baryon vertex is introduced phenomenologically 
with the coupling strength $g_\sigma$ as a free parameter and two 
choices for the mass and  the width of the $\sigma$ meson, 
$m_\sigma=\Gamma_\sigma=0.6$~GeV and $m_\sigma=\Gamma_\sigma=0.5$~GeV.
Labeling the channels by $\alpha,\beta,\gamma$, the 
Lippmann-Schwinger equation for the meson amplitude $\chi_{\alpha\gamma}$
for the process $\gamma\to\alpha$ can be cast in the form:
$$
  {\chi}_{\alpha\gamma}(k_\alpha,k_\gamma) 
         = {\mathcal{K}_{\alpha\gamma}}(k_\alpha,k_\gamma)
+ \sum_\beta\int{\rm d}k\;
  {{\mathcal{K}_{\alpha\beta}}(k_\alpha,k){\chi}_{\beta\gamma}(k,k_\gamma)
  \over \omega(k) + E_{\beta}(k)-W}\,.
$$
Approximating the kernel $\mathcal{K}$ by a separable form,
the integral equation reduces to a system of linear equations
which can be solved exactly.
For sufficiently strong coupling $g_\sigma$ the kernel  $\mathcal{K}$
may become singular and a (quasi) bound state arises.
\begin{figure}[htbp]
\begin{minipage}{63mm}
\begin{center}
\hspace{-4mm}\includegraphics[width=62mm]{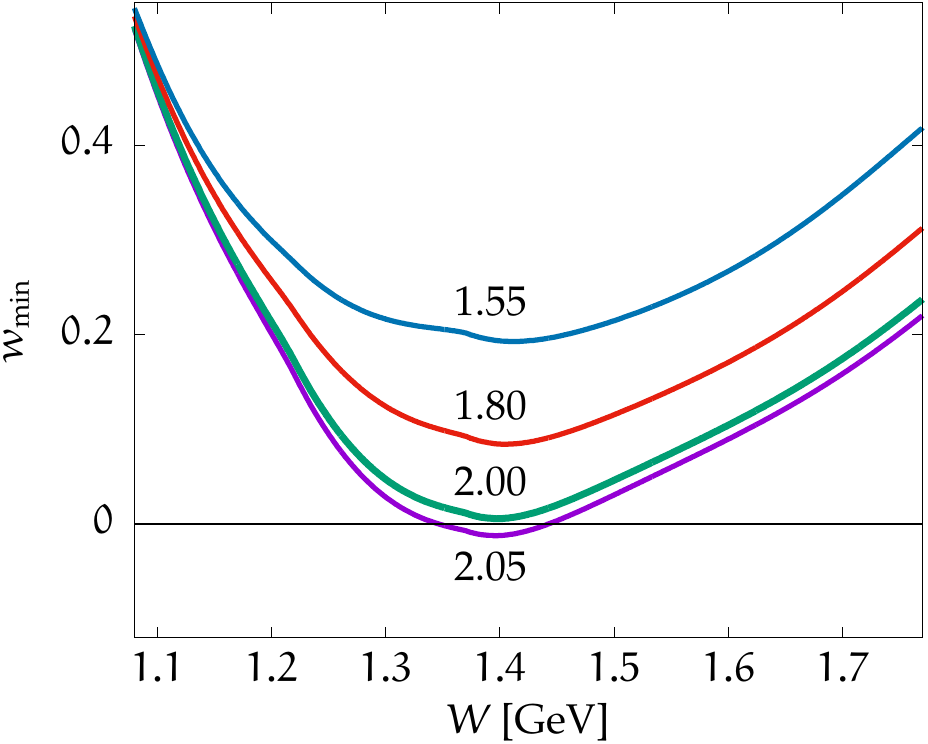}
\caption{\label{fig:wmin}
The lowest eigenvalue $w_\mathrm{min}$ for four different values
of the $\sigma N$ coupling.}
\end{center}
\end{minipage}
\hfill
\begin{minipage}{55mm}
\smallskip
{\renewcommand{\arraystretch}{1.3} 
\setlength{\tabcolsep}{8pt}
\begin{center}
\begin{tabular}{|c|c|c|}
\hline

\hline
$g_\sigma$ & Re$W_p$ & $-2{\rm Im}W_p$\\ 
         & [GeV] & [GeV] \\  
\hline

\hline
PDG & 1.370  & 0.175 \\ 
\hline

\hline
1.80 & 1.397 &  0.157 \\ 
\hline
1.95 & 1.383 &  0.112 \\ 
\hline
2.00 & 1.358 &  0.111 \\ 
\hline
2.05 & 1.331 &  0.044 \\ 
     & 1.438 &  0.147 \\ 
\hline

\hline
\end{tabular}
\end{center}
}

\vspace{6pt}

{\bf Table 1.} Poles in the complex $W$-plane
for four typical values of $g_\sigma$.
The PDG values are from \cite{PDG}.
\end{minipage}
\end{figure}

In order to study this process we follow the evolution of the lowest 
eigenvalue of the matrix pertinent to the system  of linear equations, 
$w_\mathrm{min}$, as a function of $W$ for different values of $g_\sigma$ 
(see Fig.~\ref{fig:wmin}).
Along with this evolution we observe the evolution of the resonance 
$S$-matrix pole in the complex $W$-plane using the Laurent-Pietarinen 
expansion~\cite{L+P2013,L+P2014,L+P2015,L+P2014a} (see Table 1).
We see that the lowest eigenvalue indeed touches the zero line 
for $g_\sigma=2.0$, the pole, however, emerges already for considerably 
weaker couplings and starts approaching the real axis. 
Beyond the critical value, $w_\mathrm{min}$ crosses zero twice,
producing two poles in the complex energy plane.
It is interesting to note that for the values below the critical 
value, the real part of the pole position almost coincides with $W$
at which $w_\mathrm{min}$ reaches its minimum.
This value of $W$ is of the order of 100~MeV below the nominal
$\sigma N$ threshold.
The result agrees well with the molecular picture of the Roper
resonance discussed in the first part of this contribution.
Let us note that because the $\sigma N$ channel is coupled to 
other channels, the molecular state has a finite width (i.e.
finite Im$W_p$) even for $g_s$ greater than the critical value. 

In the present approach we have also studied the influence of
including a genuine three quark state with one quark excited 
to the $2s$ orbit.
Using $g_\sigma\approx 1.5$, the results for the position as well as 
the modulus and the phase come close to the PDG value~\cite{PDG},
and are rather insensitive to the mass of the genuine three-quark 
state.
This leads us to the conclusion that the mass of the $S$-matrix
pole is determined by the energy of the molecular state while
its detailed properties may still considerably depend on the
three-quark excited state.
The simple model discussed in the first paper provides a 
simplified picture which enables a deeper insight into 
the mechanism of the resonance formation, hindered by the
complex formalism of the coupled-channel approach.

\end{document}